\pdfoutput=1
\documentclass[10pt,aps,prl,floatfix,superscriptaddress,nofootinbib,nobibnotes,
amsmath,amssymb,twocolumn]{revtex4-2}
\usepackage{graphicx}
\usepackage{xcolor}
\usepackage{mathtools}
\usepackage[colorlinks=true,citecolor=blue,linkcolor=blue,urlcolor=blue,
    breaklinks=true,hypertexnames=false]{hyperref}
\usepackage{orcidlink}

\newcommand{\bea}{\begin{eqnarray}}
\newcommand{\eea}{\end{eqnarray}}

\begin{document}

\title{
Circumstellar Medium of Supernovae as New Probes for Feebly Interacting Particles}

\author{Yu Cheng
\orcidlink{0000-0002-4822-3890}}
\email[Contact author: ]{chengyu@kaist.ac.kr}
\affiliation{Department of Physics, Korea Advanced Institute of Science and Technology (KAIST), Daejeon 34141, Korea}

\author{Chui-Fan Kong
\orcidlink{0009-0007-7010-5085}}
\email[Contact author: ]{kongcf@ibs.re.kr}
\affiliation{Particle Theory and Cosmology Group (PTC), Center for Theoretical Physics of the Universe (CTPU),
Institute for Basic Science, Daejeon 34126, Korea}

\author{Yen-Hsun Lin
\orcidlink{0000-0001-7911-7591}}
\email[Contact author: ]{yenhsun@phys.ncku.edu.tw}
\affiliation{Institute of Physics, Academia Sinica, Taipei 115, Taiwan}

\author{Meng-Ru Wu
\orcidlink{0000-0003-4960-8706}}
\email[Contact author: ]{mwu@as.edu.tw}
\affiliation{Institute of Physics, Academia Sinica, Taipei 115, Taiwan}
\affiliation{Institute of Astronomy and Astrophysics, Academia Sinica, Taipei 106, Taiwan}
\affiliation{Physics Division, National Center for Theoretical Sciences, Taipei 106, Taiwan}

\author{Seokhoon Yun
\orcidlink{0000-0002-7960-3933}}
\email[Contact author: ]{seokhoon.yun@knu.ac.kr}
\affiliation{Department of Physics, Kyungpook National University, Daegu 41566, Korea}
\affiliation{Particle Theory and Cosmology Group (PTC), Center for Theoretical Physics of the Universe (CTPU),
Institute for Basic Science, Daejeon 34126, Korea}

\begin{abstract}
We propose a novel strategy to probe feebly interacting particles (FIPs) by exploiting the dense, confined circumstellar medium (CSM) surrounding core-collapse supernovae.
FIPs produced in the protoneutron star can deposit substantial visible energy into the
CSM via decay prior to the shock breakout from the progenitor star.
This energy injection heats and ionizes the CSM, establishing an FIP-induced photosphere that generates distinctive precursor blackbody emission.
Using early-time observations of supernova~2023ixf, we translate the nondetection of excessive precursor luminosity into stringent new constraints on MeV-scale dark photons
as an exemplary model.
Our results significantly extend existing core-collapse supernova bounds and exclude previously unexplored regions of parameter space.
We further demonstrate that the FIP-induced dust sublimation offers
key
diagnostics for future Galactic supernovae, opening a new avenue to explore the dark sector.
\end{abstract}

\maketitle

\textit{Introduction}---Core-collapse supernovae provide promising environments to probe new feebly interacting particles (FIPs), such as axions, axionlike particles, dark photons, and sterile neutrinos,
complementing laboratory searches.
The protoneutron star (PNS) reaches temperatures of
$\mathcal{O} (30)\,{\rm MeV}$
and nuclear-scale
densities, enabling efficient FIP production through its interaction with the standard model particles.
The produced FIPs can free-stream out and decay or become trapped and diffuse for various supernova (SN) observables to probe their production and transport~\cite{Raffelt:1987yt,Turner:1987by,Mayle:1987as,Raffelt:1990yz,Janka:1995ir,Raffelt:1996wa,Farzan:2002wx,Raffelt:2011nc,Dent:2012mx,Kazanas:2014mca,Rrapaj:2015wgs,Chang:2016ntp,Hardy:2016kme,Fischer:2016cyd,Arguelles:2016uwb,
Jaeckel:2017tud,Chang:2018rso,DeRocco:2019njg,Sung:2019xie,Carenza:2019pxu,Bar:2019ifz,Mastrototaro:2019vug,Suliga:2019bsq,Dev:2020eam,Lucente:2020whw,Carenza:2020cis,Suliga:2020vpz,Caputo:2021rux,Fischer:2021jfm,Sung:2021swd,Calore:2021lih,Caputo:2022mah,Shin:2022ulh,Hoof:2022xbe,Ferreira:2022xlw,Lella:2022uwi,Ho:2022oaw,Diamond:2023scc,Ray:2023gtu,Chauhan:2023sci,Carenza:2023old,Muller:2023pip,Lella:2023bfb,Akita:2023iwq,DelaTorreLuque:2024zsr,Ray:2024jeu,Manzari:2024jns,Mori:2024vrf,Fiorillo:2024upk,Syvolap:2024hdh,Li:2024pcp,Lella:2024dmx,Springmann:2024ret,Vogl:2024ack,Alonso-Gonzalez:2024ems,Benabou:2024jlj,Hardy:2024gwy,Takata:2025lyu,Caputo:2025aac,Fiorillo:2025yzf,Chauhan:2025mnn,Cappiello:2025tws,Balaji:2025alr,Fiorillo:2025sln,Fiorillo:2025gnd,Candon:2025ypl,Candon:2025sdm,Mori:2025cqf,Caputo:2025avc,Mori:2025eit,Ferreira:2025qui,Gupta:2025ygk,Huang:2025xvo,Blinov:2025aha,Chauhan:2025xqr,Joseph:2026nut,Caputo:2026pdw,Yu:2026xtb}.

The canonical constraint is the SN~1987A cooling bound~\cite{Raffelt:1990yz,Raffelt:1996wa}:
effective FIP emission from the PNS would
shorten or distort the observed neutrino burst, excluding the parameter space where the FIP luminosity competes with the neutrino luminosity at around $1\,{\rm s}$ after the core bounce.
Beyond the cooling argument, other SN probes can be competitive or even stronger in different parameter spaces.
Excessive energy deposition through FIP decay or absorption into the stellar medium outside the PNS can be constrained by
SN explosion energy~\cite{Sung:2019xie,Caputo:2022mah}.
Visible decays outside the star's photosphere can
yield prompt $\gamma$-ray signals
(constrained by the nondetection from SN~1987A~\cite{Oberauer:1993yr}), positron injection
(leading to Galactic x-ray flux), and the integrated radiative emission from cosmic core-collapse supernovae that contribute to the diffuse photon background~\cite{Dar:1986wb,Kazanas:2014mca,Jaeckel:2017tud,DeRocco:2019njg,Calore:2021lih,Balaji:2025alr}.

However, no studies have considered the impact of FIPs on the common presence of the circumstellar medium (CSM) surrounding most supernovae
produced by various mass-loss mechanisms from the progenitor stars prior to explosions~\cite{Smith:2014txa,Chandra:2017aev,Smith_2017,Fraser:2020,Dessart:2024mop}, except for a brief discussion in Ref.~\cite{Kozyreva:2024ksv}.
The interaction of the SN shock or ejecta with the surrounding CSM leaves imprints on SN light curve evolution and spectral features, which in turn, offers a critical diagnostic of the CSM properties.
In particular, recent flash spectroscopy surveys indicate that a significant fraction
of type II supernovae with red supergiant (RSG) progenitors
contains confined dense and slowly moving CSM near the progenitor, signaling enhanced progenitor mass-loss events
shortly before explosions~\cite{Forster:2018mib,Bruch:2020jcr,Jacobson-Galan:2021pki,Bruch:2022aqd}.

Recently, observations of the nearby SN~2023ixf in Messier 101 (NGC 5457)~\cite{Itagaki,Jacobson-Galan:2023ohh,Bostroem:2023dvn,Teja:2023hcm,Hiramatsu:2023inb,Zimmerman:2023mls} have provided a benchmark for this phenomenon.
Flash spectroscopy and multiwavelength modeling of SN~2023ixf have revealed that the SN~2023ixf CSM can be modeled by a two-component structure: an inner dense, confined region embedded within an outer, more standard wind~\cite{Jacobson-Galan:2023ohh,Hiramatsu:2023inb,Zimmerman:2023mls,Jacobson-Galan:2025rbd}.
Assuming the density profile of each component follows a typical spherically symmetric steady-state wind model, $\rho(r) = \dot{M} / (4\pi r^2 v_w)$, where $\dot{M}$ is the mass-loss rate and $v_w\sim 30-50$\,km\,s$^{-1}$ is the typical RSG wind velocity~\cite{1994ApJ...420..268C},
the inner and dense CSM ($\lesssim 2\times 10^{14}$\,cm) was associated with an enhanced mass-loss
($\dot{M} \sim 10^{-2}\,M_\odot\,{\rm yr^{-1}}$) shortly before the explosion, while the outer wind corresponds to mass ejection of $\dot{M} \sim 10^{-4}\,M_\odot\,{\rm yr^{-1}}$.
Notably, SN~2023ixf also exhibited significant circumstellar dust at $\sim 10^{15}$\,cm that can reprocess the progenitor's photospheric blackbody (BB) emission from optical to infrared~\cite{Kilpatrick:2023pse,Niu:2023tmz,Jencson:2023bxz}.

\begin{figure}[tbp]
\centering
\includegraphics[width=.9\columnwidth]{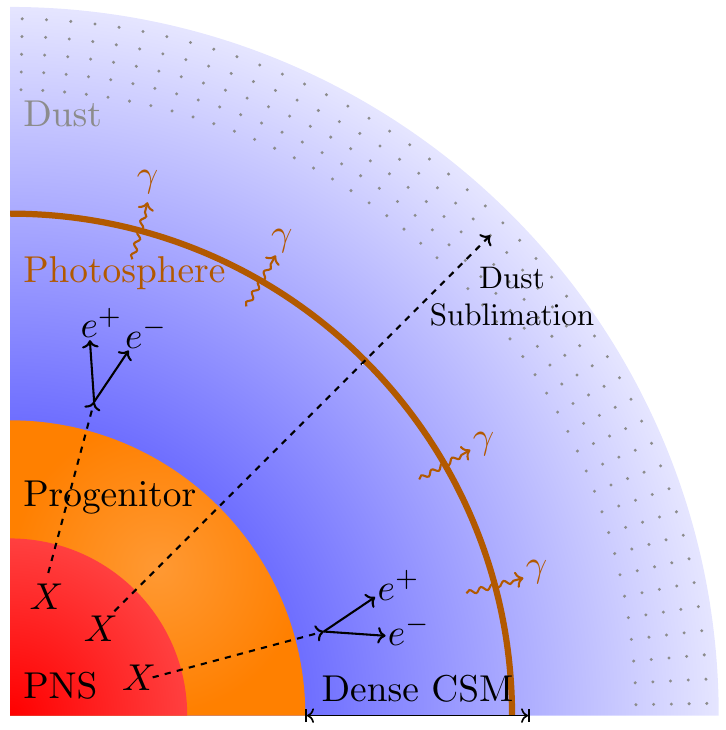}
\caption{
\label{fig:scheme}
Schematic plot showing FIPs ($X$) from the PNS depositing energy
into the CSM, forming a new photosphere in dense CSM.
The same FIP heating also leads to dust sublimation at the outer CSM, preventing obscuration of the light from the FIP-induced photosphere.
}
\end{figure}

In this \emph{Letter}, we propose a novel method to probe FIPs produced from the PNS using CSM-related observables.
The energy deposition from visible FIP decays in the CSM ahead of the shock not only raises the gas temperature and modifies the opacity, but also leads to dust sublimation.
This induces a new photosphere within the dense CSM, whose
emissions are free from dust reprocessing.
A stringent constraint on the FIP parameter space is derived using the reported upper limits for SN~2023ixf prior to shock breakout (SBO).
A schematic plot of this process is shown
in Fig.~\ref{fig:scheme}.

For the rest of this \emph{Letter}, we specifically consider the dark photon (DP) scenario, in which an additional Abelian gauge boson, denoted by $\gamma^\prime$,
interacts with the standard model through kinetic mixing with the ordinary photon by $\mathcal{L} = (\varepsilon/2)F_{\mu\nu}F^{\prime \mu\nu}$ in natural unit, where $F_{\mu\nu}^{(\prime)} = \partial_\mu A_\nu^{(\prime)}-\partial_\nu A_\mu^{(\prime)}$, with the primed and unprimed quantities referring to the DP and photon fields, respectively, and $\varepsilon$ is the dimensionless kinetic mixing parameter~\cite{Holdom:1985ag,Okun:1982xi,Galison:1983pa,Arkani-Hamed:2008hhe}.
The ingredients needed for our study, including DP production in the PNS, in-medium effects, and decay,
are summarized in Ref.~\cite{Caputo:2025avc} and references therein.
In what follows, we adopt the PNS profile obtained from the
SFHo-18.8
spherically symmetric
core-collapse supernova (CCSN) simulation~\cite{Bollig:2020phc,Garching}.
For completeness, the dependence of the results
on different CCSN simulation profiles
is provided in Supplemental Material (SM)~\cite{SupplementalMaterial}.
Hereafter, we adopt the centimeter-gram-second system of units.

\textit{DP energy deposition in CSM}---For a DP produced in PNS with momentum $k$, its decay length reads
\bea
L_d = \frac{k}{m_{\gamma'}} \Gamma_{\gamma'}^{-1} ,
\label{eq:Decaylength}
\eea
where $m_{\gamma^\prime}$ is the DP mass, and $\Gamma_{\gamma'}$ is the DP decay rate in the rest frame.
Since DPs emitted from the PNS propagate approximately radially through the CSM, the differential visible energy injection
per unit radius reads
\bea\label{eq:dQ/dr}
    \frac{d Q}{d r}
    =
    \int d k
    \frac{1}{L_d}
    \frac{d \mathcal{N}_{\gamma'}}{ d k} e^{-r/L_d}
    \sum_i \int dE_i E_i \frac{d N_i}{d E_i},
\label{eq:dQ/dt/dr/dEe}
\eea

where $d \mathcal{N}_{\gamma'}/ d k$ is the DP spectrum integrated over the production volume around the PNS size and over the emission period (set by the PNS cooling timescale $\sim 10\,\mathrm{s}$), and the sum runs over all visible final states $i=e^\pm$.
The $e^\pm$ energy spectrum per DP decay is given by
\begin{equation}\label{eq:main:dN/dEe}
    \frac{d N_e}{d E_e}
    =
    \frac{1}{ 2 \gamma v p_e} \Theta(E_e-E_-)\,\Theta(E_+-E_e)\,,
\end{equation}
where $E_\pm=\gamma( m_{\gamma'}c^2/2 \pm v p_e)$, $v$ is the DP velocity in the lab frame with $\gamma$ the associated Lorentz factor, $p_e$ is the momentum of $e^\pm$, and $\Theta$ is the Heaviside step function; see SM~\cite{SupplementalMaterial} for detail.
Hereafter, we focus on the regime where $m_{\gamma'}>2m_e$ and subdominant channels other than $\gamma'\rightarrow e^\pm$ that produce radiation, e.g., $\gamma'\to3\gamma$ \cite{McDermott:2017qcg} and $\gamma'\to e^\pm \gamma$ \cite{DeRocco:2019njg}, in the final states are neglected due to their suppressed branching ratios.
Furthermore, when $m_{\gamma'} \gtrsim 2m_\mu$, $\gamma'\to \mu^\pm$ is possible. But it is hardly attained in the interested parameter space discussed later and is also ignored.

For the typical CSM condition, $e^\pm$ from DP decay transfer their kinetic energy to the medium through scattering during their propagation.
For $e^\pm$ produced at radius $r$ with initial energy $E_e$, we compute the radial distance that they can travel in any given CSM density profiles before completely losing their kinetic energy, defined as the stopping length $\delta r_{\rm stop}(E_e,r)$, using the stopping power table
from ESTAR~\cite{estar} (see SM~\cite{SupplementalMaterial} for details).
Comparing $\delta r_{\rm stop}(E_e,r)$ to the scale radius $r$, we find that typically $\delta r_{\rm stop}(E_e,r)/r
\lesssim 0.1$ in the inner region where CSM densities are higher, i.e., $e^\pm$ efficiently deposit their energy into local CSM after production.
At outer radii where CSM density decreases, $\delta r_{\rm stop}(E_e,r)/r$ increases steeply from 0.1 to much larger values, indicating inefficient local energy deposition.
Detailed comparison of $\delta r_{\rm stop}(E_e,r)/r$ and further discussions are given in SM~\cite{SupplementalMaterial}.
Motivated by this, we
determine the local energy deposition efficiency $\eta(E_e,r)$ by comparing $\delta r_{\rm stop}(E_e,r)$ to $r$,
\begin{equation}\label{eq:eta_local}
\eta(E_e, r) =
\begin{cases}
1-(m_e c^2)/E_e, & \delta r_{\rm stop}(E_e,r) \leq r, \\
\delta E_e/E_e, & \delta r_{\rm stop}(E_e,r) > r,
\end{cases}
\end{equation}
where $\delta E_e$ is the $e^\pm$ energy loss after traveling over the scale radius $r$.
Note that this approximation mostly underestimates the true energy deposition at regions where local energy deposition is inefficient (see SM~\cite{SupplementalMaterial}), and can be considered as a conservative approach.

\begin{figure}[tbp]
\centering
\includegraphics[width=\columnwidth]{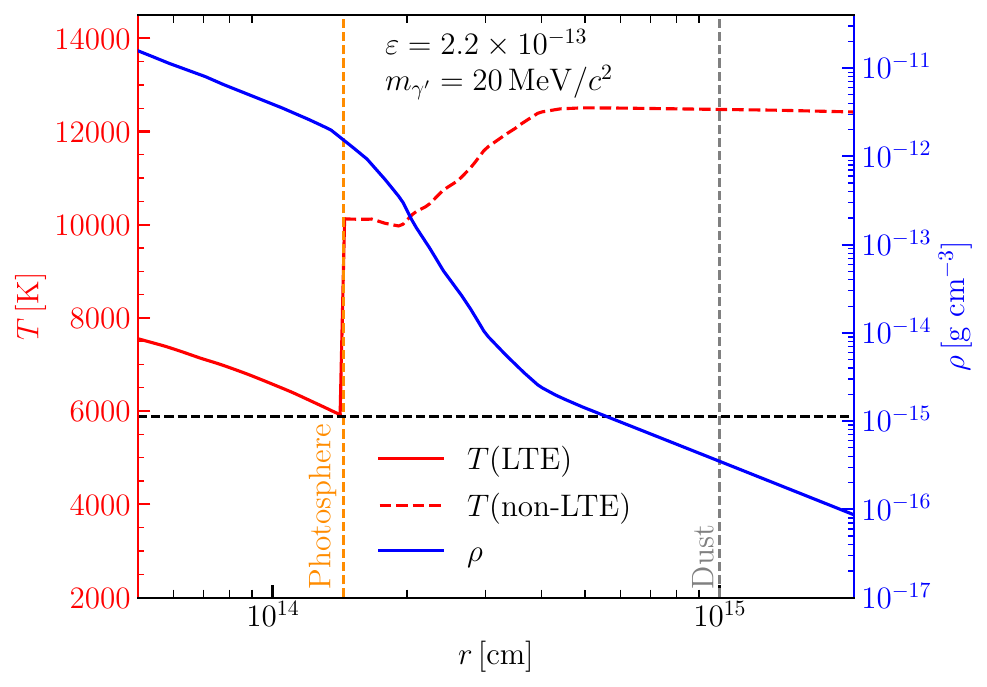}
\caption{\label{fig:DensityTprofile}
CSM density profile from Ref.~\cite{Zimmerman:2023mls} (blue line)
features a dense region for $r \lesssim 2\times 10^{14}$\,cm followed by a transition to the outer dilute region.
The red line shows the temperature profile following DP energy deposition for the benchmark $(\varepsilon, m_{\gamma'})$. The dashed orange line denotes the DP-induced photosphere at $r_{\rm ph}\simeq 1.4\times 10^{14}$\,cm
and $T_{\rm ph}\simeq 5800$\,K.
Estimates show that the CSM reaches the temperature profile shown by the red line in less than $\sim 1000$\,s  (see EM).
}
\end{figure}

Since the injected $e^\pm$ energy spectra are
determined by the DP spectrum and decay kinematics, we define the spectral-averaged deposition efficiency,
\begin{equation}\label{eq:eta_bar}
\bar{\eta}(r)=\frac{\int dE_e \frac{dQ(r)}{dr dE_e}  \eta(E_e,r)}{ \int dE_e \frac{dQ(r)}{dr dE_e} }.
\end{equation}
Taking the SN~2023ixf CSM density profile from Ref.~\cite{Zimmerman:2023mls}, shown by the solid blue line in Fig.~\ref{fig:DensityTprofile}, and assuming the CSM as an ideal gas
with a solarlike mass composition of $70\%$ hydrogen, $28\%$ helium, and $2\%$ heavier elements (consisting of $0.6\%$ carbon, $0.6\%$ oxygen, $0.4\%$ nitrogen, and $0.4\%$ iron),
we find $\bar{\eta} \simeq 1$ for $r \lesssim 2\times 10^{14}\,{\rm cm}$ for relevant DP parameter space explored here.
Beyond this radius, $\bar{\eta}$ drops significantly below 0.1 as the CSM density decreases; see, e.g., Fig.~\ref{fig:eta_tdep} in End Matter (EM).
The timescale for DP energy deposition can be approximated similarly to Eq.~\eqref{eq:eta_bar} and can be used to estimate the DP heating rate $\dot q_{\rm heat}$.
We find that in the region where $\bar\eta\simeq 1$, the energy deposition time scale ($\bar t_{\rm dep}\lesssim 1000$\,s) is much shorter than the dynamical timescale $t_{\rm dyn} \sim r/c_s = \mathcal{O}(1{\rm -}10)\,{\rm yr}$, where $c_s$ is the local sound speed.
Energy deposition from DPs therefore can be considered as instantaneous in inner dense CSM.
At outer part, $\bar t_{\rm dep}$ can be substantially longer.
A detailed evaluation of $\bar\eta$, $\dot q_{\rm heat}$, and their comparison with the local radiative cooling rate $|\dot q_{\rm cool}|$ is provided in EM.

\textit{DP-induced CSM luminosity}---Energy deposition from
DP decay can heat the initially cold CSM gas of
$\sim 1000$~K~\cite{Dessart:2017pfi,Jacobson-Galan:2021pki} before SBO.
To determine the postheating gas temperature, it is crucial to
examine whether
radiation is in local thermal equilibrium (LTE) or not.
The key distinction between the LTE and non-LTE regimes is the role of trapped radiation.
In the LTE regime, photons are efficiently trapped within the medium, so their internal energy must be included when determining the thermal response of the gas.
Because the radiation energy density grows steeply with temperature, this substantially increases the
heat capacity and suppresses the rise of the gas temperature.
In contrast, in the non-LTE regime, radiation decouples from the medium and
does not contribute to the heat capacity.
As a result, the deposited energy can heat the gas more effectively.

Taking the very uncertain initial CSM temperature as zero as a conservative approach, we first estimate the postheating
gas temperature $T$ assuming LTE by solving
\begin{equation}\label{eq:dQ_tot}
\frac{dQ}{dr}\frac{\bar\eta}{4\pi r^2}=
u_{\rm gas} + I_{\rm ion} + u_{\rm rad},
\end{equation}
where
$u_{\rm gas}$ and $u_{\rm rad}$ are the internal energy density of ideal gas and radiation, respectively, and $I_{\rm ion}$ is the ionization energy density.
We take $u_{\rm gas} = (3/2)(n + n_e)k_B T$, where $n$ is the total number density of neutral atoms and ions, $n_e=\sum_i n_i x_i$ is the free-electron density with $x_i$ the ionization fraction for atomic species $i$ determined by solving the Saha equation~\cite{1920Natur.105..232S,1921RSPSA..99..135S}, and $k_B$ is the Boltzmann constant;
$I_{\rm ion}=\sum_i n_i x_i \chi_i$ with $\chi_i$ the ionization energy of species $i$ and $u_{\rm rad} = 4\sigma_{\rm SB}T^4/c$
with $\sigma_{\rm SB}$ the Stefan-Boltzmann constant.

Next, we check the validity of the LTE condition by evaluating the Rosseland optical depth $\tau(r)=\int_{r}^\infty dr \kappa_R(r)\rho(r)$, where the Rosseland mean opacity $\kappa_R$ is taken from OPAL~\cite{OPAL1996} and Ferguson~\cite{Ferguson:2005pu} tables for the high ($T \geq 10^4\,{\rm K}$) and low temperature ($T < 10^4\,{\rm K}$) regimes, respectively, with solarlike composition and the obtained $T$ profile assuming LTE.
The LTE solution is then only applied to $r<r_{\rm ph}$, where $r_{\rm ph}$ is the DP-induced new photosphere radius with $\tau(r_{\rm ph})=2/3$.
The solid red curve in Fig.~\ref{fig:DensityTprofile} shows $T(r)$ in the LTE regime for a benchmark DP parameter set $(\varepsilon, m_{\gamma^\prime}) = (2.2 \times 10^{-13}, 20\,{\rm MeV}/c^2)$.
The DP-induced photosphere at $r_{\rm ph}\simeq 1.4\times 10^{14}\,{\rm cm}$ with the temperature $T_{\rm ph}\approx 5800\,{\rm K}$ is marked by the vertical orange dashed line.
It is worth pointing out that
$T_{\rm ph}\approx 5800$\,K is tightly related to the temperature dependence of $\kappa_R$.
As $T$ increases over 5000\,K,
$\kappa_R$ experiences a sharp increase due to
${\rm H}^-$ formation~\cite{1939ApJ....90..611W,1945ApJ...102..223C}, a prerequisite leading to efficient photon trapping in dense CSM.
Beyond dense CSM ($r\gtrsim 2\times 10^{14}$\,cm), even if
a larger $\varepsilon$ can heat the region above
$5800\,{\rm K}$, the low density cannot lead to photon trapping.
Our numerical calculations indicate that $r_{\rm ph}$ in $(\varepsilon,m_{\gamma'})$ parameter space consistently locates at
$\sim(1{\rm-}3)\times10^{14}$\,cm, with
$T_{\rm ph}$ larger than $\sim 5800\,{\rm K}$.

Outside $r_{\rm ph}$ where LTE no longer holds, the energy deposition can also heat up the CSM but the radiative cooling needs to be taken into account, whose rate ($\dot q_{\rm cool}$) varies substantially with local gas temperature \cite{Schure:2009bx}.
We determine the non-LTE gas temperature by taking the lower value from two temperatures obtained by satisfying $\dot q_{\rm heat}=|\dot q_{\rm cool}(T)|$ and by solving Eq.~\eqref{eq:dQ_tot} without including $I_{\rm ion}$ and $u_{\rm rad}$.
The dashed red line outside $r_{\rm ph}$ shows the corresponding solution
for the benchmark $(\varepsilon,m_{\gamma'})$.
The values span between $10\,000$ and $13\,000$\,K.
We also note that even in the LTE regime, $\dot q_{\rm heat}$ must exceed $|\dot q_{\rm cool}|$ before the gas temperature surpasses $5800\,{\rm K}$ to trap the radiation; see EM for discussion.

The DP-heated CSM exemplified by Fig.~\ref{fig:DensityTprofile} implies a distinct transient emission phase, whose properties qualitatively differ from both the pre-SN stage and the later SBO phase.
The elevated gas temperature outside $r_{\rm ph}$ naturally destroys preexisting CSM dust (likely carbonaceous and silicate grains~\cite{Kilpatrick:2023pse, Jencson:2023bxz} with a characteristic sublimation temperature $T_{\rm sub} \sim 2000\,{\rm K}$).
As a result, radiation emerging from the newly established photosphere at $r_{\rm ph}$ is expected to be unaffected by dust reprocessing and to resemble a near-BB spectrum with $T_{\rm ph}\approx 5800$~K.
The corresponding BB luminosity reads
\begin{equation}\label{eq:LBB}
L_{\rm BB}\simeq 4\pi r_{\rm ph}^2 \sigma_{\rm SB} T_{\rm ph}^4,
\end{equation}
which amounts to $\sim 1.6\times 10^{40}\,\rm erg\,s^{-1}$ for the benchmark profile in Fig.~\ref{fig:DensityTprofile}.
We find an emission timescale of $\sim 20$ h after the arrival of DPs in the CSM (see EM).
This prediction can be confronted with early-time data for SN~2023ixf to derive the constraints presented below.

\begin{figure}[tbp]
\centering
\includegraphics[width=\columnwidth]{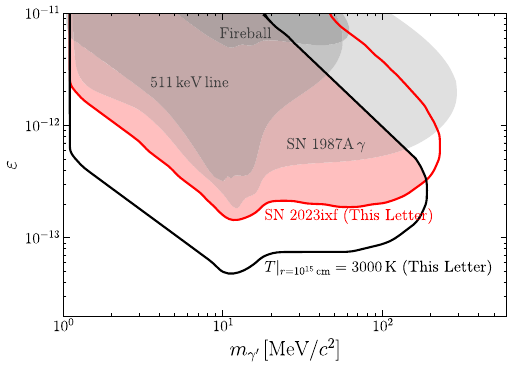}
\caption{
The solid red contour indicates the parameter that violates
Eq.~\eqref{eq:L_BB_constraint} and
the solid black curve marks $T|_{r=10^{15}\,{\rm cm}}>3000$\,K.
The red-shaded region is
excluded by the considerations adopted in this \emph{Letter}.
Gray-shaded regions show existing CCSN-associated constraints evaluated using the SFHo-18.8 profile.
\label{fig:all}
}
\end{figure}

\textit{Constraint from SN~2023ixf}---SN~2023ixf was first reported by K.~Itagaki via the Transient Name Server~\cite{Itagaki}, and the earliest detection was later confirmed at MJD 60082.85 (20:29 UTC on May 18, 2023).
We use extensive archival data, including
(non)detections from amateur astronomers~\cite{amateur_TNS}, to place the onset of core collapse at approximately 9--25
h before the first detection (see EM for the discussion on the uncertainty of this timing from the progenitor size and shock velocity).
We utilize three subsequent nondetections within a 10-h window (see Fig.~\ref{fig:amaeteur_photometry}) following the inferred
core-collapse time to set a conservative upper limit on the DP-induced BB luminosity,
\begin{equation}\label{eq:L_BB_constraint}
\iota(T_{\rm ph}) L_{\rm BB} \leq 8 \times 10^{39}\,{\rm erg\,s}^{-1}\,,
\end{equation}
where $\iota(T_{\rm ph})\simeq 0.54$ accounts for instrument-specific bandwidth sensitivities.
This threshold is chosen to accommodate the heterogeneity of amateur telescope configurations and potential systematic uncertainties in the CSM modeling.
Further details on the bandwidth-corrected luminosity estimates are provided in EM.

The DP parameter space that violates Eq.~\eqref{eq:L_BB_constraint} is represented by the solid red contour in Fig.~\ref{fig:all}.
Along this contour, the corresponding photospheric temperature is $T_{\rm ph}\simeq 5800\,{\rm K}$ due to the rapidly rising opacity discussed above, which in turn localizes the photosphere radius to $r_{\rm ph}\simeq 1.4\times 10^{14}\,{\rm cm}$, consistent with Fig.~\ref{fig:DensityTprofile}.
Since the energy density at $T_{\rm ph}$ is dominated by radiation (i.e., $u_{\rm rad}\gg u_{\rm gas}\,,I_{\rm ion}$), one can estimate the upper limit on total DP energy deposited within $r_{\rm ph}$ as $\sim (4/3)\pi r_{\rm ph}^3\times u_{\rm rad}(T_{\rm ph})\sim\mathcal{O}(10^{44})\,{\rm erg}$, close to the threshold found numerically,
\begin{equation}\label{eq:Q_th}
Q_{\rm th} \leq 2.7
\times 10^{44}\,{\rm erg}.
\end{equation}
While the DP-heated CSM temperature profile along the lower part of the red contour shares a similar profile as shown in Fig.~\ref{fig:DensityTprofile}, the upper contour requires further discussion.
For larger $\varepsilon$ and $m_{\gamma^\prime}$, where the typical DP decay length becomes smaller than the dense CSM radius, the DP energy deposition profile $dQ/dr$ falls off rapidly for $r\gtrsim r_{\rm ph}$.
Consequently, the outer CSM is only weakly heated, so that dust at $r\sim 10^{15}\,{\rm cm}$ may survive, and the BB emission can be reprocessed into the infrared, weakening the optical constraint.
We therefore impose an additional dust-destruction requirement, $T|_{r=10^{15}\,\rm{cm}}>3000\,{\rm K}$, shown as the solid black contour in  Fig.~\ref{fig:all}.
The overlap (red-shaded) region, where the luminosity bound (solid red) intersects this dust-destruction criterion, is
hence
excluded by considering both effects.
In the remaining region above the black curve but below the red curve, a dedicated treatment of dust sublimation and radiative reprocessing is required for a robust conclusion.

A few comments are in order for the modeling uncertainties associated with Fig.~\ref{fig:all}.
The major one comes from the underlying CCSN model that affects the DP production.
Taking a different CCSN model can lead to up to $\sim 50\%$ variation in $\varepsilon$ for $m_{\gamma'}\lesssim 150$~MeV (see SM~\cite{SupplementalMaterial} for details).
Importantly, nearly the same variation applies to other bounds (gray-shaded regions) so that the overall improvement from the CSM consideration remains unaffected.
We also verified that adopting an alternative best-fit CSM density profile from Ref.~\cite{Jacobson-Galan:2023ohh} results in nearly identical exclusion contours (see also SM~\cite{SupplementalMaterial}).
Moreover, taking a more conservative choice of 4000~K for dust sublimation only results in less than $10\%$ changes to the black contour.

CSM dust sublimation driven by DP heating may
provide a powerful diagnostic for the next Galactic (or nearby) CCSN associated with an RSG progenitor, for which CSM dust typically resides at $r\sim 10^{15}$\,cm~\cite{1994AJ....107.1469D,Verhoelst:2009vv}.
We find that the contour corresponding to $T|_{r=10^{15}\,\rm{cm}}>3000\,{\rm K}$ is insensitive to the
conventional CSM wind profile with $\dot M\sim 10^{-6}-10^{-4}\,M_\odot\,{\rm yr}^{-1}$.
This robustness against variations in $\dot{M}$ and $v_w$ arises from a cancellation in the energy balance in Eq.~\eqref{eq:dQ_tot}.
In the low-density regime ($\bar{\eta} \ll 1$), the deposition efficiency scales as $\bar{\eta} \propto \delta r_{\rm stop} \propto \rho \propto \dot{M}/v_w$, while the gas internal energy density
follows the same scaling, $u_{\rm gas} \propto n \propto \rho \propto \dot{M}/v_w$.
Therefore, in the non-LTE region and neglecting ionization, the required $dQ/dr$ to heat the gas at a given radius to a fixed temperature is approximately independent of $\dot M/v_w$.
With low-latency electromagnetic follow-up observation triggered by the neutrino alerts from a nearby SN~\cite{SNEWS:2020tbu}, a rapid spectral transition from infrared-excess to optical or UV dominance between core collapse and SBO would be a strong indication of dust sublimation by exotic energy injection.
Conversely, the absence of such a transition would further strengthen the CSM-based bound on DP as shown by Fig.~\ref{fig:all}.

\textit{Summary and discussion}---We have shown that the CSM surrounding CCSN progenitors and its related observables offer a new avenue to probe FIPs.
By considering the DP scenario and SN~2023ixf, we have demonstrated that visible energy deposition from DP decay into $e^\pm$ pairs can create an optically thick region in a dense CSM, enclosed by a DP-induced photosphere with a temperature $\gtrsim 5800$\,K, and can additionally sublimate dust in the outer CSM.
Together, these effects result in a transient optical signal
between core-collapse time and SBO.
Exploiting early-time nondetections for SN~2023ixf, we have derived an extended DP constraint beyond the SN~1987A $\gamma$ bound for $2 m_e<m_{\gamma'}\leq 200$\,MeV$/c^2$, and quantified the corresponding upper limit on DP energy deposition in the dense CSM of SN~2023ixf [Eq.~\eqref{eq:Q_th}].
Our results further indicate that the next nearby SN with a RSG progenitor could provide enhanced sensitivity through CSM dust sublimation signatures.
We note that the method developed here is generally applicable to other FIP candidates.

We caution that the results presented here are based on specific astrophysical scenarios adopted to deduce the CCSN and CSM properties associated with SN~2023ixf, including the inference of the explosion time and the assumption of sphericity.
While we do not expect qualitative changes when considering more complicated SBO models or aspherical CSM profiles, further investigations along these lines should be pursued.

The presence of CSM may impact other existing FIP bounds, particularly those relying on visible decays outside stellar progenitors.
Other signatures beyond the simple BB emission discussed in this \emph{Letter}, including the detailed spectral line features related to the change of CSM ionization and recombination emission, are well motivated but require dedicated non-LTE radiative-transfer calculations. Finally, nearby RSGs such as Betelgeuse may provide an attractive test bed for probing FIP-induced dust destruction in a well-characterized stellar environment.
We plan to explore all these exciting directions in future work.

\bigskip
\textit{Acknowledgments}---We are grateful to Tobias Fischer and Thomas Janka for granting us the use of their SN models.
Y.\ C. is supported by the National Research Foundation of Korea (NRF) Grant No.~RS-2023-00211732, by the Samsung Science and Technology Foundation under Project No.~SSTF-BA2302-05, by the POSCO Science Fellowship of POSCO TJ Park Foundation, and by the NRF Grant No.~RS-2024-00405629.
C.-F.\ K. and S.\ Y. are supported by IBS under the Project code No.~IBS-R018-D1.
S.\ Y. is also supported by the Basic Science Research Program through the NRF funded by the Ministry of Education (No.~RS-2026-25571203).
Y.-H.\ L. and M.-R.\ W. acknowledge support of the National Science and Technology Council, Taiwan under Grants No.~115-2112-M-001-037-MY3, No.~111-2628-M-001-003-MY4, No.~115-2112-M-001-054-MY5, and Academia Sinica under Project No.~AS-IV-114-M04.
Y.-H.\ L. further acknowledges support from the Shui-Chin Lee Foundation through the Shui-Chin Lee Fellowship.
M.-R.\ W. also acknowledges support of the Physics Division of the National Center for Theoretical Sciences, Taiwan.
C.-F.\ K. and S.\ Y. thank APCTP, Pohang, Korea for their hospitality during the Focus Program [No.~APCTP-2026-F01] from which this work greatly benefited.

\clearpage

\appendix
\onecolumngrid
\section*{End Matter}
\twocolumngrid

\section{$e^\pm$ energy deposition timescale}\label{sec:tdep}
\setcounter{equation}{0}
\renewcommand{\theequation}{A\arabic{equation}}

\begin{figure}[h]
\centering
\includegraphics[width=0.9\columnwidth]{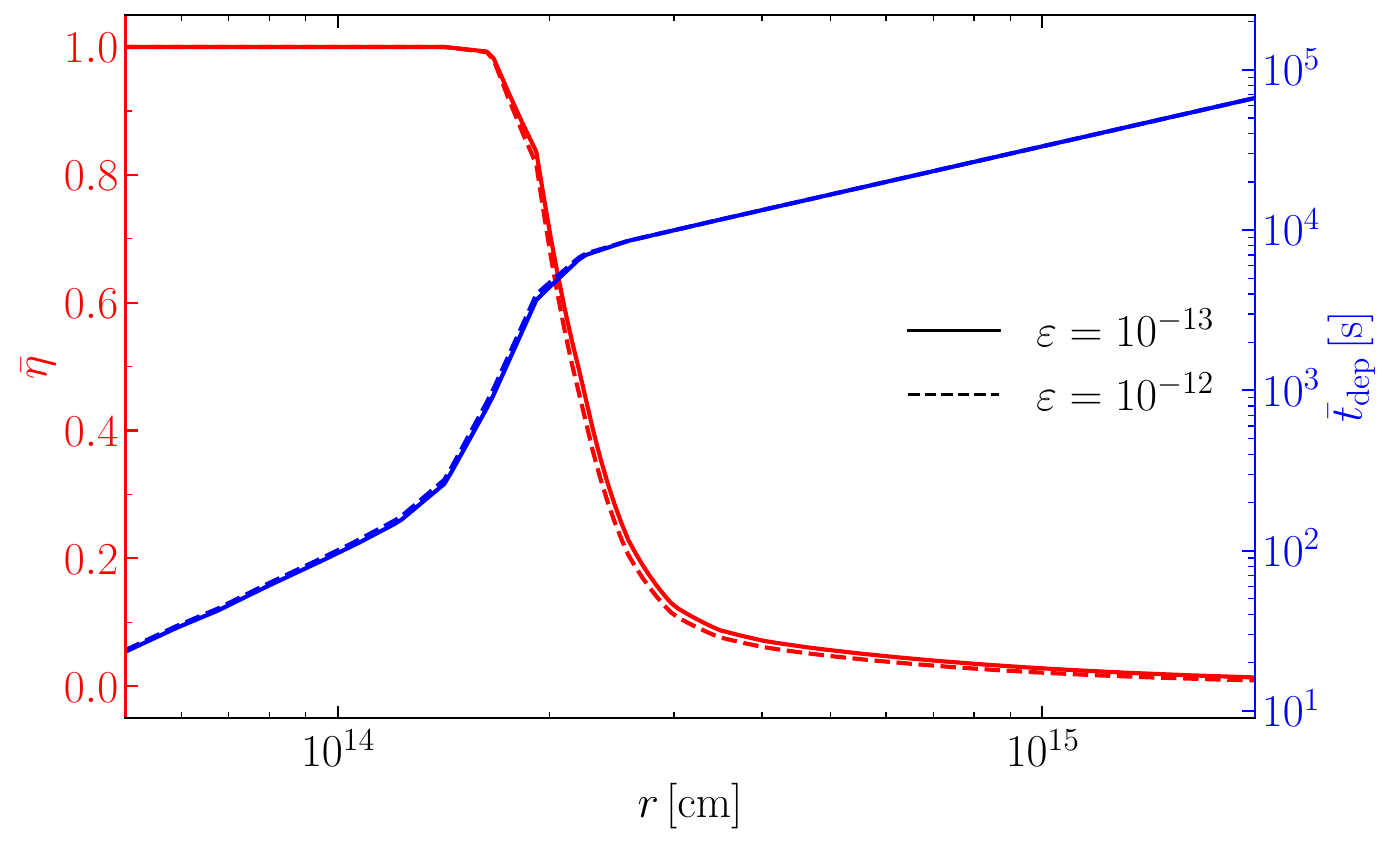}
\caption{
The spectral-averaged energy deposition efficiency $\bar{\eta}$ (red) and energy deposition time $\bar{t}_{\rm dep}$ (blue) as functions of $r$ for CSM density profile
shown in Fig.~\ref{fig:DensityTprofile}.
Two different values of $\varepsilon=10^{-13}$ (solid) and $\varepsilon=10^{-12}$ (dashed) are illustrated.
Both assume $m_{\gamma'}=10$\,MeV$/c^2$.
\label{fig:eta_tdep}
}
\end{figure}

The characteristic timescale for an $e^\pm$ to deposit energy locally can be estimated by the stopping length,  $t \sim \delta r_{\rm stop}/c$, as the energy deposition mainly takes place when $e^\pm$ is relativistic.
Following the same approach for computing
$\eta(E_e,r)$ in the main text, we define the energy deposition timescale as $t_{\rm dep}(E_e, r) = \min(\delta r_{\rm stop}, \Delta r)/c$.
The spectral-averaged value reads
\begin{equation}\label{eq:t_bar}
\bar{t}_{\rm dep}(r)
=\frac{\int dE_e \frac{dQ(r)}{dr dE_e} t_{\rm dep}(E_e,r)}{\int dE_e \frac{dQ(r)}{dr dE_e}}.
\end{equation}
We show both $\bar{\eta}$ and $\bar{t}_{\rm dep}$ as functions of $r$ for $\varepsilon=10^{-13}$, $10^{-12}$, and $m_{\gamma'}=10$\,MeV$/c^2$ in Fig.~\ref{fig:eta_tdep}, calculated using Eqs.~\eqref{eq:eta_bar} and \eqref{eq:t_bar}, respectively.
Note that both quantities are only mildly sensitive to the choice of $(\varepsilon\,, m_{\gamma'})$.

\section{Heating and cooling rates and various timescales}\label{sec:qdot}

\begin{figure}
\centering
\includegraphics[width=.9\columnwidth]{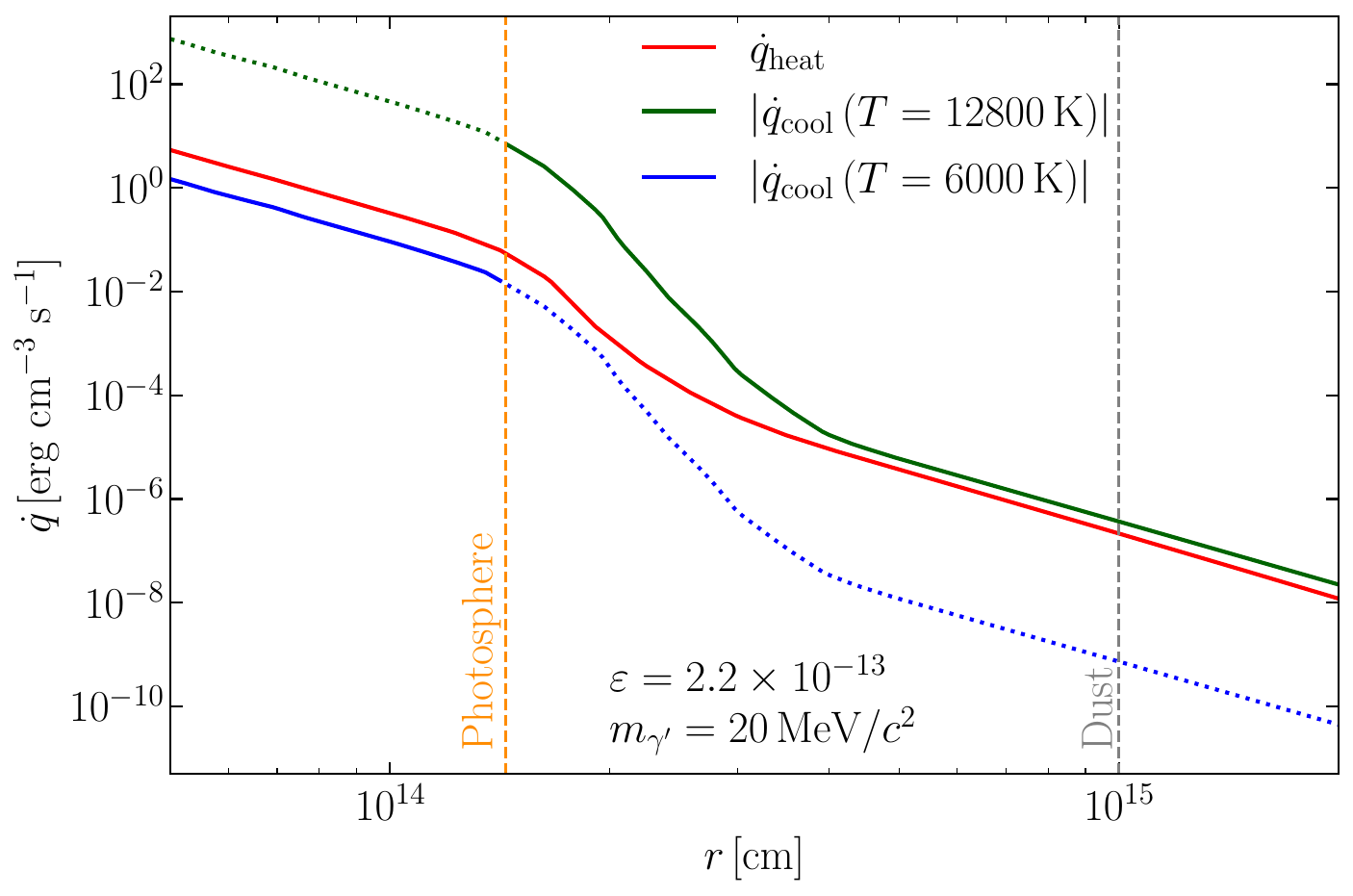}
\caption{
DP-heating rate $\dot{q}_{\rm heat}$ (red) for benchmark $(\varepsilon, m_{\gamma'}) = (2.2\times 10^{-13}\,,20\,{\rm MeV}/c^2)$ and the radiative cooling rates $|\dot{q}_{\rm cool}|$ for $T=12800$ (green) and 6000\,K (blue); see text for details.
\label{fig:qdot}
}
\end{figure}

Figure~\ref{fig:qdot} compares the DP heating rate, $\dot{q}_{\rm heat}=(\bar{\eta}/{\bar{t}_{\rm dep}}) dQ/dr$ (red), with the radiative cooling rate, $|\dot{q}_{\rm cool}|=n_{\rm H}^2\Lambda(T)$, as functions of radius $r$ for benchmark $(\varepsilon, m_{\gamma'})= (2.2 \times 10^{-13}\,, 20\,{\rm MeV}/c^2)$.
Here, $\Lambda(T)$ is the radiative loss function normalized to $n_{\rm H}$~\cite{Schure:2009bx}.
We evaluate $|\dot{q}_{\rm cool}|$ at two representative temperatures: $6000\,{\rm K}$ (blue) and $12800\,{\rm K}$ (green); solid lines indicate the regions in which each choice is relevant.

At $r<r_{\rm ph}$, radiative cooling is only relevant before photon trapping occurs at $T\gtrsim 5800\,{\rm K}$ (see main text).
We adopt $6000\,{\rm K}$ as a conservative value for the maximal radiative cooling rate during the DP-heating phase.
Even with this choice, we find that $\dot q_{\rm heat}$ exceeds $|\dot q_{\rm cool}|$
by at least a factor of 2.

For $r>r_{\rm ph}$, the gas temperature is determined by the balance condition $\dot{q}_{\rm heat}=|\dot{q}_{\rm cool}|$; we denote the corresponding equilibrium temperature by $T_{\rm eq}$, which is approximately
$12800\,{\rm K}$ for $r\gtrsim 4\times 10^{14}\,{\rm cm}$.
We notice that for substantially larger $\varepsilon$, $\dot q_{\rm heat}$ can become so large that the radiative cooling cannot balance the heating.
In that regime, the gas continues to be heated until $\dot{q}_{\rm heat}$ terminates, and the temperature can rise well above $\mathcal{O}(10^6\,{\rm K})$.

The DP-heating and radiative-cooling rates discussed above can be used to estimate several characteristic timescales relevant to our analysis.
First, the injected energy density required to elevate the gas temperature to $3000\,{\rm K}$ at $r = 10^{15}\,{\rm cm}$ is $ u_{\rm gas} \simeq 10^{-4}\,{\rm erg\,cm^{-3}}$.
At the same radius, the benchmark $(\varepsilon, m_{\gamma'})$ yields $\dot{q}_{\rm heat}\simeq 4.1\times 10^{-7}\,{\rm erg\,cm^{-3}\,s^{-1}}$.
The corresponding timescale to sublimate the dust, $t_{\rm sub}\sim u_{\rm gas}/\dot{q}_{\rm heat} \simeq 243\,{\rm s}\ll \bar{t}_{\rm dep}$, is much shorter than the deposition duration $\bar{t}_{\rm dep}$ (see Fig.~\ref{fig:eta_tdep}).
The dust is thus destroyed rapidly, providing a substantial time window for the emission from the DP-induced photosphere to be constrained by Eq.~\eqref{eq:L_BB_constraint}.
We also numerically verify that $t_{\rm sub}$ remains comparable along the lower red contour bounded by the solid black curve
in Fig.~\ref{fig:all}.

At the same location, even after dust sublimation, DPs continue to deposit energy and heat the gas further until the temperature reaches $T_{\rm eq}\simeq 12800\,{\rm K}$, which only takes an additional $\sim 800\,{\rm s}$.
After reaching $T_{\rm eq}$, the gas remains at $T_{\rm eq}$ for $\bar{t}_{\rm dep} \simeq 9\,{\rm h}$.
A naive cooling-time estimate, $u_{\rm gas}(T_{\rm eq})/|\dot{q}_{\rm cool}(T_{\rm eq})|$, gives $\sim 0.3\,{\rm hr}$; however, because $|\dot{q}_{\rm cool}|$ decreases as $T$ drops, cooling from $T_{\rm eq}$ down to $T_{\rm sub}$ actually takes substantially longer.
An estimation shows that to cool from $12800$ to 6000\,K, 6000 to 4000\,K, 4000 to 3000\,K, and 3000\,K to $T_{\rm sub}$ takes 30, 25, 14, and 16\,h, respectively, approximately 85\,h in total.
Furthermore, even after the gas cools down to $T_{\rm sub}$, the dust reformation
requires additional time.
As a result, we conclude that dust does not reform
to reobscure the DP-heated photospheric emission over the time window relevant for comparison with the early nondetections.

Another key timescale is the radiative diffusion time through the optically thick region inside $r_{\rm ph}$, which approximately sets the duration of the DP-induced photospheric emission.
For the benchmark case in Fig.~\ref{fig:DensityTprofile}, the average optical depth inside $r_{\rm ph}$ is $\bar{\tau}\simeq 17$, and it takes similar or larger values across the DP parameter space enclosed by the red contour in Fig.~\ref{fig:all}.
The DP-induced photosphere emission duration is therefore expected to be $\bar{\tau} r_{\rm ph}/c\gtrsim 20$\,h.

\section{Relevant early-time observational data}
\setcounter{equation}{0}
\renewcommand{\theequation}{C\arabic{equation}}

\begin{figure}
\centering
\includegraphics[width=1.\columnwidth]{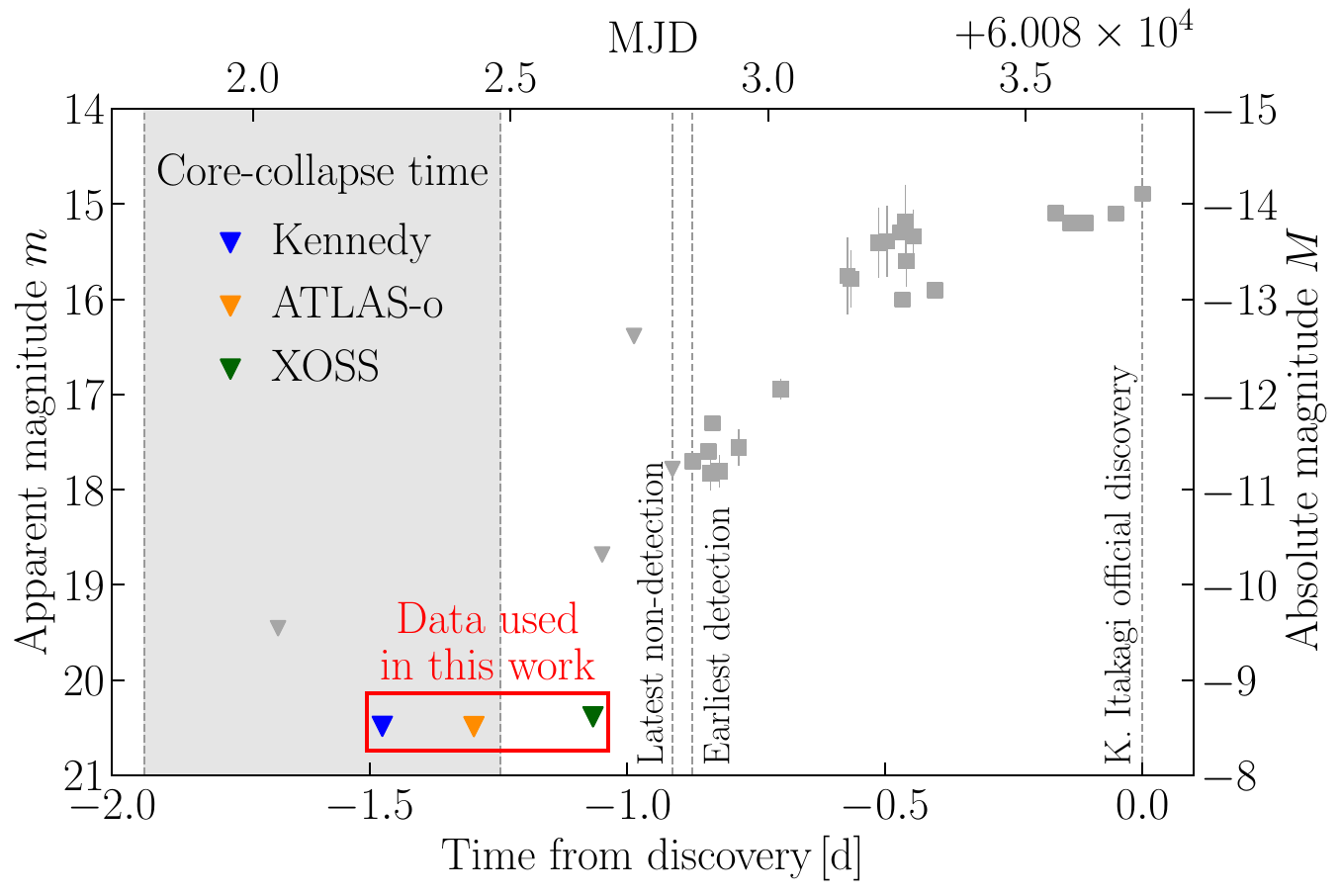}
\caption{Nondetection upper limits (triangles) and early-time detections (squares) of SN~2023ixf.
The gray band indicates the estimated time interval when the core collapse occurs.
Colored triangles represent the data used to derive
constraints in this \emph{Letter}. Figure
adapted from Ref.~\cite{Yaron_others}.
\label{fig:amaeteur_photometry}
}
\end{figure}

Using the SN~2023ixf amateur photometry from the Transient Name Server~\cite{amateur_TNS} (where $0\,{\rm d}$ marks K.~Itagaki's report \cite{Itagaki}; Fig.~\ref{fig:amaeteur_photometry}), we bypass the complex modeling of early brightness evolution \cite{Li:2023vux,Hosseinzadeh:2023ixa,Kozyreva:2024ksv}. Instead, we utilize the three nondetection upper limits (colored triangles) to constrain the DP heating.

Assuming the SN SBO
from the progenitor stellar radius
$r_s \simeq 5 \times 10^{13}\,{\rm cm}$~\cite{Zimmerman:2023mls}
occurs at $\simeq -0.93\,{\rm d}$ (MJD 60082.788) \cite{Li:2023vux}
and taking a representative
shock velocity $v_s \approx 8000\,{\rm km\,s^{-1}}$ \cite{Zimmerman:2023mls} as the averaged shock speed during its propagation inside the progenitor,
the corresponding SN core collapse can be estimated to happen
at $-1.64\,{\rm d}$ ($ r_s/v_s \approx 17\,{\rm h}$) prior to detection.
While it is difficult to pinpoint the true core-collapse time, we consider the dominating uncertainties due to the variation of shock velocities and progenitor size quoted in literature works.
Accounting for shock velocity ranging from $(8{\rm -}10)\times 10^3$\,km\,s$^{-1}$~\cite{Jacobson-Galan:2023ohh,Zimmerman:2023mls,Dickinson:2024cab}
and the progenitor size from $(2.8{\rm-}7)\times10^{13}$\,cm~\cite{Zimmerman:2023mls,Hosseinzadeh:2023ixa,2024MNRAS.534..271Q,Bersten:2023vct},
we estimate that core-collapse time may sit anywhere inside the gray band, roughly between $-1.94$ to $-1.25$\,d, shown in Fig.~\ref{fig:amaeteur_photometry}.

Since DPs produced
after the core bounce arrive the CSM in $\sim 1$\,h and it takes $\lesssim 0.3$\,h to heat up the CSM (see previous sections),
the estimated $\sim 20$\,h DP-induced photospheric emissions can be compared to the three upper limits between $-1.56$ and $-1.1\,{\rm d}$: MJD 60082.25 (Kennedy \cite{Kennedy_TNS}), MJD 60082.43 (ATLAS-o \cite{ATLAS-o_TNS}), and MJD 60082.66 (XOSS \cite{XOSS_TNS}) to establish the constraint within this time frame.
Regarding instrumentation, Kennedy used an unfiltered Sony IMX571 APS-C sensor (350--1050\,nm), and ATLAS-o employed the ATLAS orange filter (560--820\,nm). XOSS utilized a 0.2\,m {\it f}/7 Schmidt-Cassegrain telescope with an assumed IMX571-equivalent bandwidth, as specific CMOS details were undocumented for this epoch. For all three nondetections, the band-limited absolute magnitude upper limits are $M\simeq -8.6$.

For simplicity, we adopt a representative bandpass of 400--900\,nm to derive the band-limited luminosity $L$ from the corresponding band-limited absolute magnitude $M$ by
\begin{equation}\label{eq:LMrelation}
\frac{L}{L_{\odot}'} = 10^{0.4(M_{\odot}' - M)},
\end{equation}
where
$L_{\odot}'$ and $M_{\odot}'$ are the solar luminosity and absolute magnitude in the same 400--900~nm range.
$L_{\odot}'$ and $M_{\odot}'$ can be related to the solar luminosity $L_\odot$ and absolute magnitude $M_\odot$ across the full-wavelength range by
\begin{alignat}{1}
L_{\odot}' & =\iota_\odot  L_{\odot}\approx2.03\times10^{33}\,{\rm erg\,s^{-1}},\\
M_{\odot}' & =M_{\odot}-2.5\log_{10}\iota_\odot\approx5.41,
\end{alignat}
where $\iota_\odot = \iota(T_\odot) \approx 0.54$ for $T_\odot=5772$\,K, calculated by
\begin{equation}\label{eq:iota}
\iota(T) = \frac{\int_{400\,{\rm nm}}^{900\,{\rm nm}} B_\lambda(T) d\lambda}{\int_0^\infty B_\lambda(T) d\lambda},
\end{equation}
which accounts for the fraction of the total power emitted within 400--900~nm for BB radiation with
$B_\lambda(T)$
the corresponding Planck distribution.
Using Eqs.~\eqref{eq:LMrelation}--\eqref{eq:iota} and taking $M \geq -8.6$ gives
\begin{equation}\label{eq:L_DP}
L \leq 8.16 \times 10^{38}\,{\rm erg\,s^{-1}}.
\end{equation}
To remain conservative and account for potential uncertainties in amateur photometry as well as other minor factors from galaxy dust extinction and red-shift corrections, we disfavor any DP parameters that produce a luminosity within 400--900~nm exceeding
\begin{equation}
\iota(T_{\rm ph})L_{\rm BB}(\varepsilon, m_{\gamma'}) \leq 8 \times 10^{39}\, {\rm erg\,s^{-1}},
\end{equation}
which is roughly a factor of 10 larger
than Eq.~\eqref{eq:L_DP} and represents a more conservative limits.

\clearpage
\onecolumngrid

\setcounter{equation}{0}
\setcounter{figure}{0}
\setcounter{section}{0}
\setcounter{table}{0}
\setcounter{page}{1}

\renewcommand{\theequation}{S\arabic{equation}}
\renewcommand{\thefigure}{S\arabic{figure}}
\renewcommand{\thetable}{S\arabic{table}}

\begin{center}
\textbf{\large
Circumstellar Medium of Supernovae as New Probes for Feebly Interacting Particles}

\vspace{0.05in}
{ \it \large Supplemental Material}\\
\vspace{0.05in}
{Yu Cheng, Chui-Fan Kong, Yen-Hsun Lin, Meng-Ru Wu, and Seokhoon Yun}
\end{center}

\twocolumngrid
In this supplementary, we use natural unit with $\hbar=c=1$ for simplicity.

\section{$e^\pm$ energy spectrum from dark photon decay}

Considering a dark photon (DP) that has mass $m_{\gamma'}$, energy $ \omega$, and momentum $k=|\mathbf{k}|=\sqrt{\omega^2 -m_{\gamma'}^2}$, its lab-frame velocity is $v = k/ \omega$.
In the DP rest frame, the two-body decay $\gamma' \to e^\pm$ is isotropic, and each $e^\pm$ has fixed energy $ m_{\gamma'}/2$.
Let $\theta$ denote the relative angle between the $e^\pm$ momentum $\mathbf{p}_e$ in the DP rest frame and the boost direction $\mathbf{v}$.
The lab-frame $e^\pm$ energy follows from the Lorentz boost
\begin{eqnarray}
    E_{e,\theta} = \gamma \left(\frac{m_{\gamma'}}{2} + v p_e \cos \theta\right)\,,
\end{eqnarray}
where $p_e = |\mathbf{p}_e| = \sqrt{m^2_{\gamma^{\prime}}/4 - m_e^2}$ and $\gamma = \omega/m_{\gamma'}$.
Integrating over the isotropic angular distribution (i.e., $\cos\theta$ is uniformly distributed in $[-1,1]$), the lab-frame differential energy spectrum of the emitted electron/positron after is
\begin{alignat}{1}\label{eq:dN/dEe}
    \frac{d N_e}{d E_e} (\omega)
    &=
    \frac{1}{2}\int^{1}_{-1} d \cos \theta
    \delta ( E_e - E_{e,\theta}) \nonumber  \\
    &=
    \frac{1}{ 2 \gamma v p_e} \Theta(E_e-E_-)\,\Theta(E_+-E_e)\,,
\end{alignat}
where $E_\pm=\gamma( m_{\gamma'}/2 \pm v p_e)$.
The prefactor $1/2$ arises from the isotropic two-body decay in the DP rest frame.
Substituting $dN_e/ d E_e$ into $dQ/dr$ given by Eq.~\eqref{eq:dQ/dt/dr/dEe} then yields the differential $e^\pm$ luminosity deposited at radius $r$, per shell thickness $dr$ and per $e^\pm$ energy $E_e$.

\begin{figure}[h]
\centering
\includegraphics[width=\columnwidth]{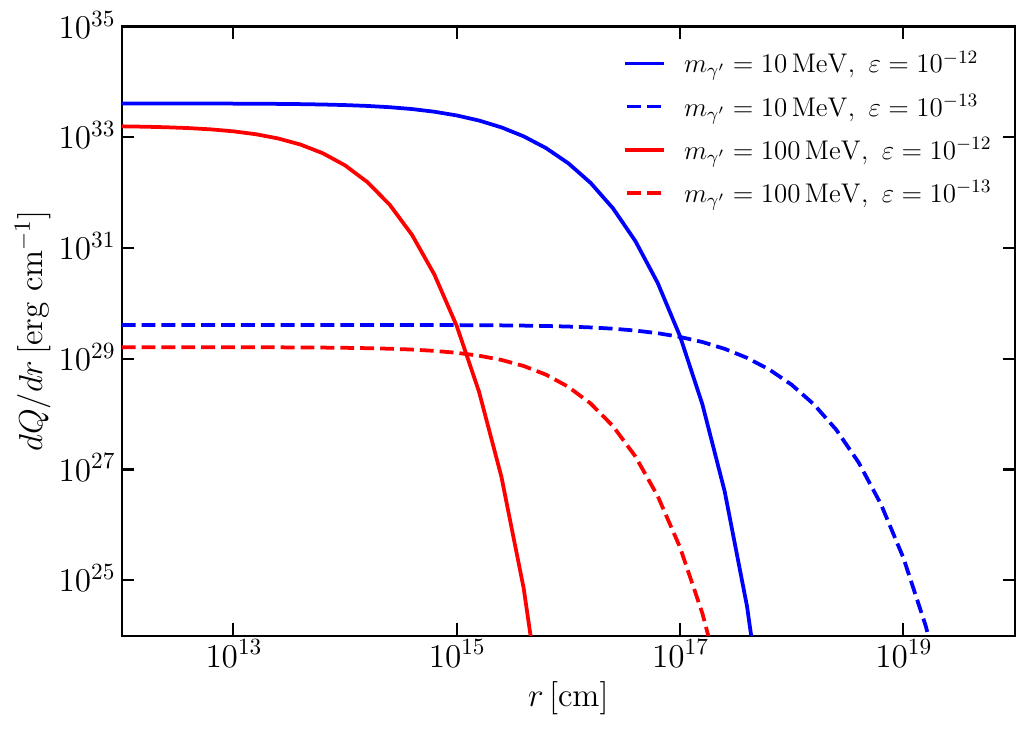}
\caption{
The differential DP visible energy injection per unit radius  $dQ/dr$ as a function of $r$ with the fiducial SN simulation model SFHo-18.8.
Four cases corresponding to $\varepsilon = 10^{-12}$ and $10^{-13}$ as well as $m_{\gamma'}=10$ and $100$\,MeV are shown.
\label{fig:dQdr}
}
\end{figure}

Using Eq.~\eqref{eq:dN/dEe}, we compute $dQ/dr$ for the SFHo-18.8 model and show it as a function of $r$ in Fig.~\ref{fig:dQdr} for representative values of $(\varepsilon\,,m_{\gamma'})$.
After integrating over DP momentum, the result is approximately constant for $r \ll \bar{L}_d$ and exhibits an exponential falloff at $r \gtrsim \bar{L}_d$ due to the exponential decay factor $e^{-r/L_d}$.
For example, the momentum-averaged DP decay length for $(\varepsilon\,,m_{\gamma'})=(10^{-13},100\,{\rm MeV})$ is $\bar L_d \simeq 4 \times 10^{15}\,{\rm cm}$.

\section{Comparison of DP constraints from different EoS models}

\begin{figure}
\centering
\includegraphics[width=\columnwidth]{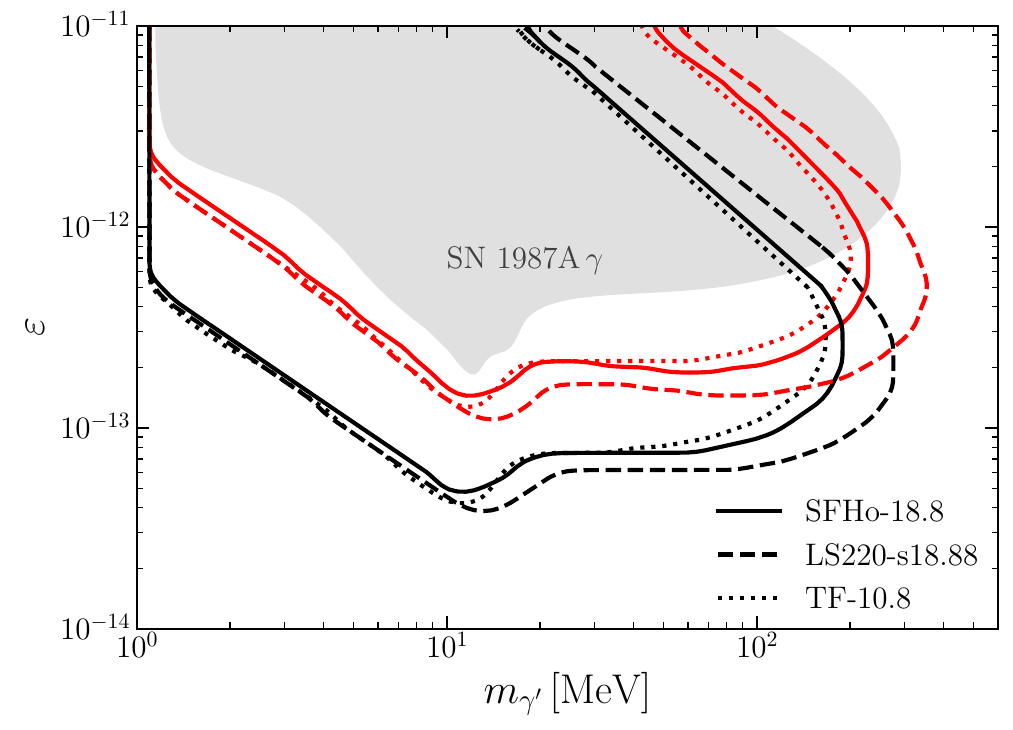}
\caption{
The derived CSM-based constraints using different SN models: SFHo-18.8 (solid line), LS220-18.88 (dashed line) and TF-10.8 (dotted line).
\label{fig:EoS_comparison}
}
\end{figure}

We investigate the sensitivity of the excluded DP parameter space
to different nuclear equation of state (EoS) and progenitor mass by considering three representative models: SFHo-18.8 and LS220-s18.88 models from Ref.~\cite{Bollig:2020phc,Garching}, and the TF-10.8 from Ref.~\cite{Fischer:2009af}.
These models primarily differ in the PNS temperature evolution, which controls the DP production rate and spectrum.

As shown in Fig.~\ref{fig:EoS_comparison}, the resulting constraints exhibit only modest model-to-model variation, with LS220-s18.88 yielding the strongest bound.
We note that the prompt SN~1987A $\gamma$ limit shown in the figure
is firstly derived using the SFHo-18.8 model while it was LS220-18.88 in Ref.~\cite{Caputo:2025avc}.
This limit (as well as other SN bounds) is therefore expected to carry a comparable degree of SN-model dependence as our CSM-based bounds.
Even without accounting for that dependence explicitly, our CSM-based constraints remain stronger than the SN~1987A $\gamma$ bound shown in Fig.~\ref{fig:EoS_comparison}, regardless of the choice of the SN model.

\section{Electron energy loss after propagation and the energy deposition efficiency}

\begin{figure}
\centering
\includegraphics[width=.9\columnwidth]{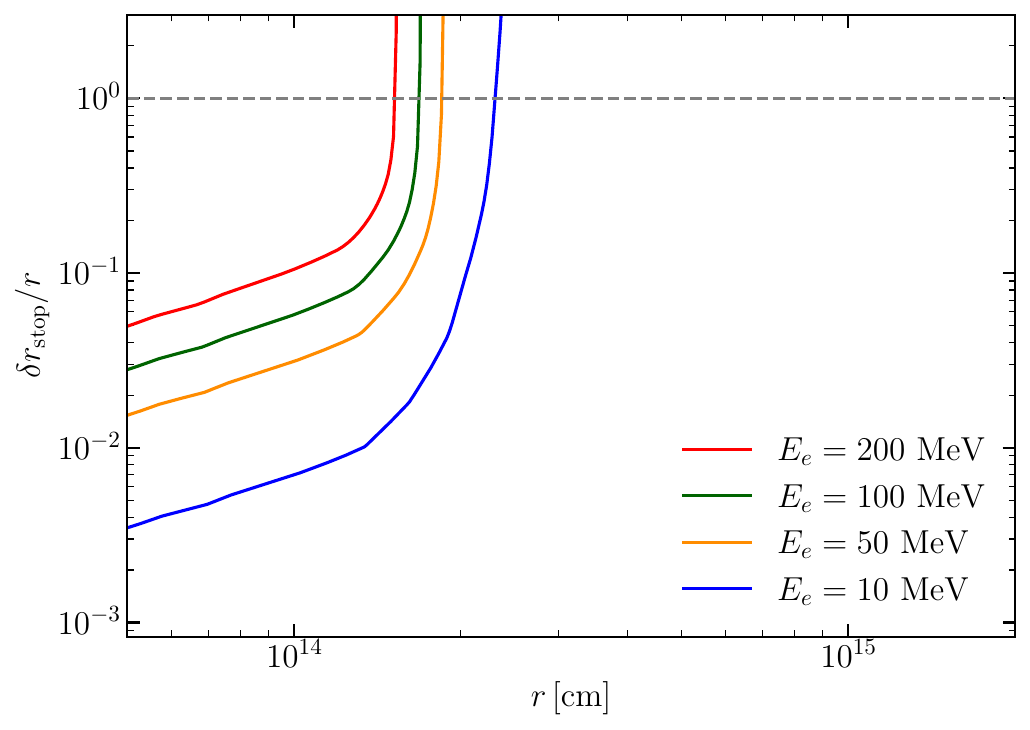}
\caption{
The ratio of $\delta r_{\rm stop}/r$ at different radii $r$ for various initial electron energies $E_e$.
In the regions $r \lesssim 1.5 \times 10^{14}$\,cm, the ratio remains below unity, indicating efficient energy deposition.
Roughly at $r\gtrsim 2\times10^{14}$\,cm, $e^\pm$ cannot be fully stopped before reaching the considered CSM boundary at $r=10^{16}\,{\rm cm}$ for all relevant $E_e$ in this work.
\label{fig:stop_ratio}
}
\end{figure}

\begin{figure}
\centering
\includegraphics[width=.9\columnwidth]{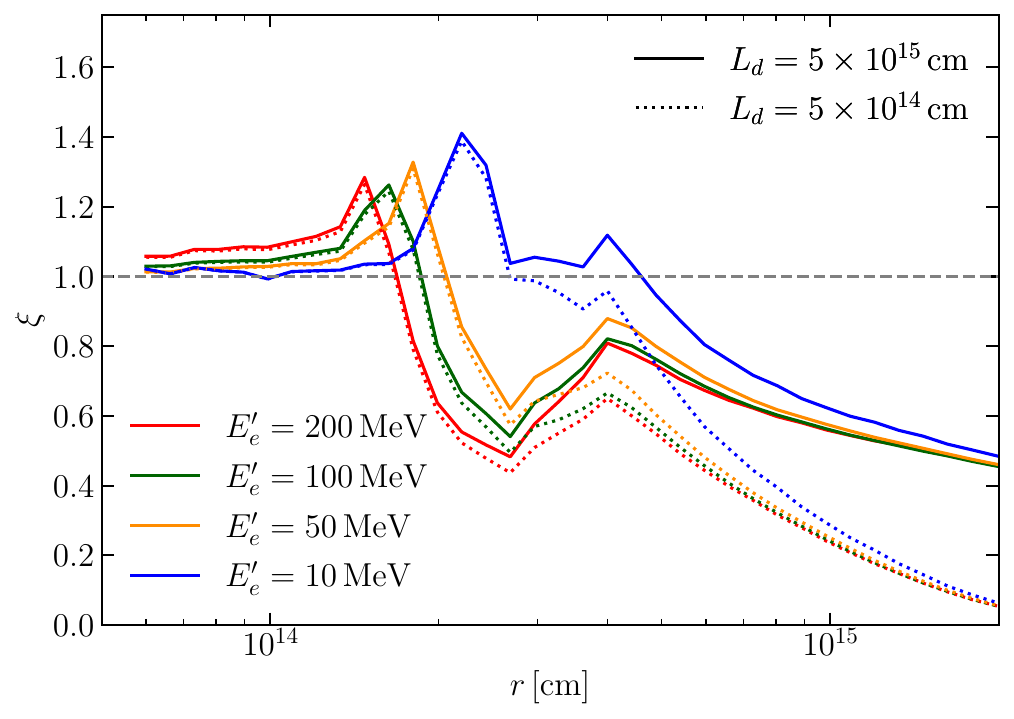}
\caption{\label{fig:nonlocal_over_local}
The quantity $\xi(r)$ that compares the local energy deposition efficiency to the actual energy deposition considering the nonlocal effect for different $E_e'$ and $L_d$ [see Eq.~\eqref{eq:xi}].
$\xi\simeq 1$ at $r<r_{\rm stop}$, validating the local energy deposition approximation.
At $r>r_{\rm stop}$, $\xi$ decreases due to the increasingly important nonlocal contributions ignored in Eq.~\eqref{eq:eta_local}.}
\end{figure}

The energy loss per unit mass thickness for an electron (or positron) of energy $E$ propagating through a medium of density $\rho(r)$ is characterized by the stopping power~\cite{estar}
\begin{equation}\label{eq:stopping_power}
\mathcal{S}(E) = - \frac{1}{\rho(r)} \frac{dE}{dr},
\end{equation}
where $dE/dr$ is the energy loss per unit path length.
For an electron with initial energy $E_e$, its remaining energy after propagating from $r_i$ to $r_f$ can be obtained by solving
the differential equation Eq.~\eqref{eq:stopping_power}.
Setting the remaining energy as $m_e$ defines the stopping length $\delta r_{\rm stop}=r_f-r_i$.

Fig.~\ref{fig:stop_ratio} shows the ratio $\delta r_{\rm stop}/r$ to the CSM radii $r$.
Across the parameter space of interest, we find that for a broad range of initial energies $E_e$, the stopping length is much smaller than the CSM radii for $r \lesssim 1.5 \times 10^{14}\,{\rm cm}$, implying that most of the $e^\pm$ energy is deposited into CSM locally.
This figure also shows that outside the radius where $\delta r_{\rm stop}\sim 0.1 r$, $\delta r_{\rm stop}/r$ increases steeply, meaning that their energy deposition extends much beyond the scale radius $r$.

In the main text, we have used Eq.~\eqref{eq:eta_local} to approximate the local energy deposition efficiency in our computation.
This treatment decouples the DP decay part [Eq.~\eqref{eq:dQ/dr}] from the the $e^{\pm}$ deposition, not only allowing for easier computation, but also providing clearer physics picture developed in the discussion of the main text.
Here, we additionally justify the validity of Eq.~\eqref{eq:eta_local}.

The full energy deposition rate per unit radius $dE_{\rm dep}(r)/dr$ without taking the local energy deposition approximation can be written as

\begin{widetext}
\bea\label{eq:fulldE/dr}
    \frac{d E_{\rm dep}(r)}{d r}
    =
    \int d k
    \frac{1}{L_d}
    \frac{d \mathcal{N}_{\gamma'}}{ d k}
    \int_{r_{\rm min}}^{r}dr'
    e^{-r'/L_d}
    \sum_i \int dE'_i \frac{d N_i}{d E'_i}\frac{dE(r,E'_i,r')}{dr},
\eea
\end{widetext}
where $E'_i$ ($i=e^\pm$) denotes the $e^\pm$ energy at $r'$ where they are produced by DP decay, $dE(r,E'_i,r')/dr$ is the same energy loss per unit path length function used in Eq.~\eqref{eq:stopping_power} evaluated at $r$ for $e^\pm$ produced at $r'$ with energy $E'_i$, and $r_{\rm min}$ is the progenitor radius.
This equation basically sums the energy deposition per unit radius at $r$ over \emph{all} $e^\pm$ that are produced inside $r$.

From Eq.~\eqref{eq:fulldE/dr}, it is clear that for given $E'_i$ and the DP decay length $L_d$, we can define
\bea\label{eq:xi}
\xi(r) \equiv  \eta(r) \bigg / \left [\frac{1}{E'_i}\int_{r_{\rm min}}^{r}dr'
    e^{-r'/L_d}
    \frac{dE(r,E'_i,r')}{dr} \right ],
\eea
to compare the deviation of the approximated local energy deposition efficiency $\eta(r)$ from the actual energy deposition.
Fig.~\ref{fig:nonlocal_over_local} shows $\xi(r)$ for several $E'_e$ and two representative $L_d$ values relevant for this study.
Comparing  Fig.~\ref{fig:nonlocal_over_local} to Fig.~\ref{fig:stop_ratio}  shows that $\eta\simeq 1$ is indeed a very good approximation in regions where $r_{\rm stop}\leq r$.
At radii where $r_{\rm stop}>r$, our choice of Eq.~\eqref{eq:eta_local} generally underestimates the actual energy deposition rate.
This is because in this limit, most of the energy deposition originates from $e^\pm$ produced at $r\lesssim L_d$ that can continuously deposit their energy far outside their production radii.
At $r= 10^{15}$~cm where we consider the dust sublimation, Eq.~\eqref{eq:eta_local} only accounts for $\sim 25\%-50\%$ of the actual energy deposition.

Overall, this comparison validates the local energy deposition assumption at inner CSM region relevant to the luminosity bound, and demonstrates that our energy deposition treatment is indeed conservative for the outer CSM region relevant for dust sublimation.

\section{Opacity}

\begin{figure}
\centering
\includegraphics[width=.9\columnwidth]{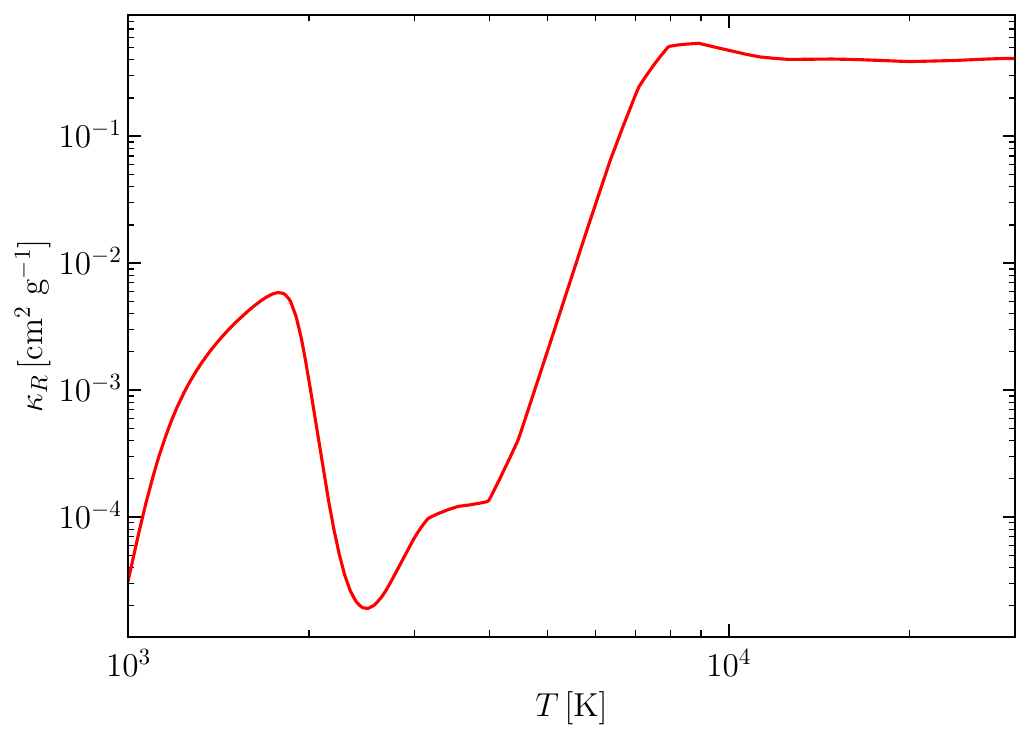}
\caption{
The Rosseland mean opacity $\kappa_R$ as a function of temperature, assuming $\rho \approx 2 \times 10^{-12}\,{\rm g\,cm^{-3}}$.
\label{fig:kappa}
}
\end{figure}

\begin{figure}
\centering
\includegraphics[height=.7\columnwidth]{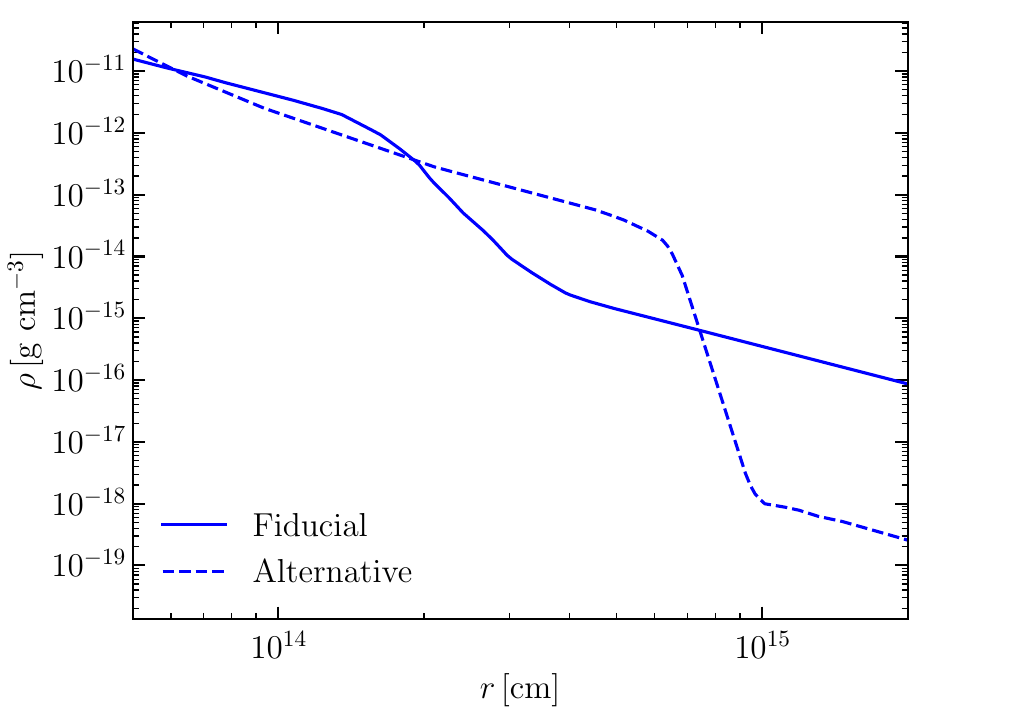}
\caption{
\label{fig:diff_profile_eta_tdep}
Comparison of the fiducial (solid) SN~2023ixf CSM density profile \cite{Zimmerman:2023mls} used in the main text and alternative one (dashed) from Ref.~\cite{Jacobson-Galan:2023ohh}.
}
\end{figure}

To evaluate the optical depth $\tau$, we adopt the Rosseland mean opacity $\kappa_R$ from the Ferguson~\cite{Ferguson:2005pu} and OPAL~\cite{OPAL1996} tables for $T < 10^4$\,K and $T \geq 10^4$\,K, respectively.
Assuming a medium density of $\rho \approx 2 \times 10^{-12}\,{\rm g\, cm^{-3}}$, which is the mean value in the dense regime, Fig.~\ref{fig:kappa} shows $\kappa_R$ as a function of temperatures.
The steep rise of $\kappa_R$ for $T \gtrsim 5000$\,K, approaching $\gtrsim 0.1\,{\rm cm}^2\,{\rm g}^{-1}$, is related to efficient ${\rm H}^-$ formation.
This increase in opacity is the main reason that photon-trapping becomes efficient in the inner CSM heated by DP decays.

\section{Comparison of two different SN 2023ixf CSM density profiles}

\begin{figure}[!htbp]
\centering
\includegraphics[width=.85\columnwidth]{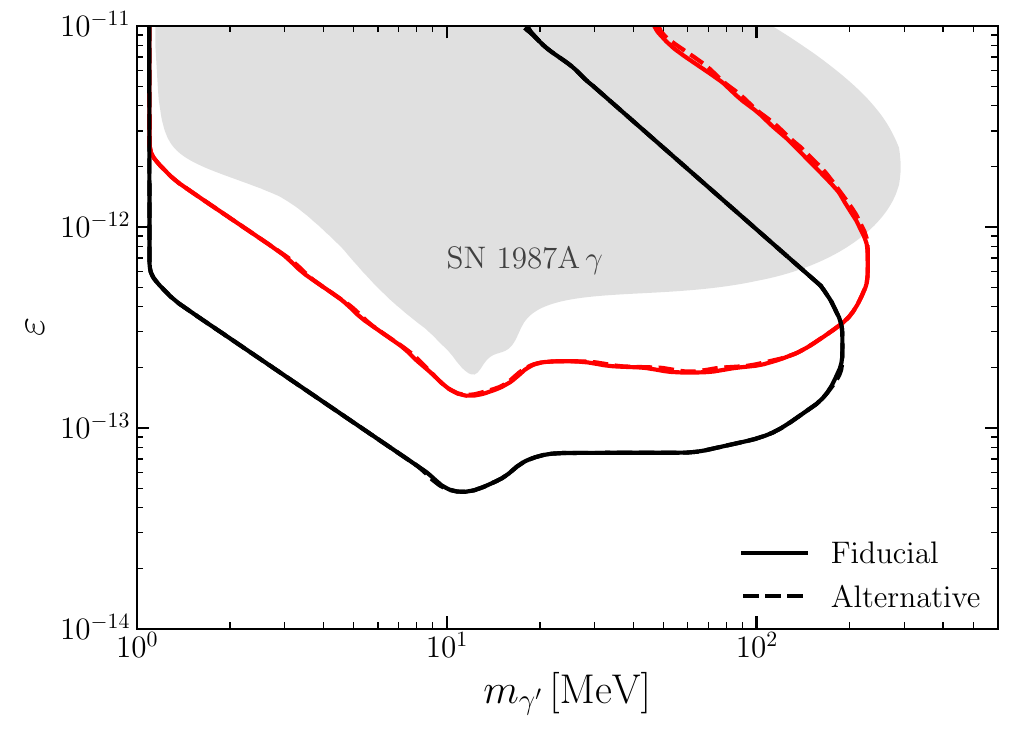}
\caption{\label{fig:diff_profile_constraint}
Same as Fig.~\ref{fig:all}, but derived using the fiducial (solid) and alternative (dashed) CSM density profiles.
}
\end{figure}

In this subsection, we explore the dependence of our result on the choice of the CSM density profile.
Fig.~\ref{fig:diff_profile_eta_tdep} compares the fiducial CSM profile from Ref.~\cite{Zimmerman:2023mls} used in the main text to another profile from Ref.~\cite{Jacobson-Galan:2023ohh}, and Fig.~\ref{fig:diff_profile_constraint} compares the corresponding contours from the luminosity constraint and dust sublimation criterion.
Despite the generally large difference in CSM profiles, particularly at $r\gtrsim 2\times 10^{14}$~cm, both contours are nearly unaffected.
For the luminosity constraint (red contours), this is because the DP-induced photospheres  locate at similar radii and have similar temperatures.
For the dust sublimation criterion (black contours), as discussed in the main text, the result is insensitive to the CSM density because both the heating and the internal energy of the gas are linearly proportional to the density $\rho$, which cancel out the density dependence.

\end{document}